\documentclass[12pt]{article}
\usepackage[margin=2cm]{geometry}
\usepackage{comment}
\usepackage{amsmath,amssymb,extarrows,mathtools,graphicx,subfigure,setspace}
\usepackage{cite}
\usepackage{slashed}
\usepackage{color}
\makeatother

\newcommand{\be}{\begin{equation}}
\newcommand{\bea}{\begin{eqnarray}}
\newcommand{\eea}{\end{eqnarray}}
\newcommand{\ba}{\begin{array}}
\newcommand{\ea}{\end{array}}
\newcommand{\ee}{\end{equation}}
\newcommand{\bes}{\begin{equation*}}
\newcommand{\beas}{\begin{eqnarray*}}
\newcommand{\eeas}{\end{eqnarray*}}
\newcommand{\bas}{\begin{array*}}
\newcommand{\eas}{\end{array*}}
\newcommand{\ees}{\end{equation*}}

\numberwithin{equation}{section}
\begin{document}
\onehalfspacing
\vfill
\begin{titlepage}
\vspace{10mm}
\begin{flushright}
 %IPM/P-2012/??? \\
%FPAUO-12/10\\
\end{flushright}
%\vbox{
 %   \halign{#\hfil
 %          IPM/P-2012/050
  %
 %                     }  end of \halign
  %    }  % end of \vbox
\vspace*{20mm}
\begin{center}
{\Large {\bf Analytical holographic superconductors in $AdS_N$-Lifshitz topological  black holes    }\\
}

\vspace*{15mm}
\vspace*{1mm}
{D. Momeni$^{a}$\footnote{e-mail:d.momeni@yahoo.com}, R. Myrzakulov $^{a}$\footnote{e-mail:r.myrzakulov@gmail.com}, L. Sebastiani $^{a}$\footnote{e-mail:l.sebastiani@science.unitn.it} and M. R. Setare$^{b}$\footnote{e-mail:rezakord@ipm.ir}}

 \vspace*{1cm}

{\it ${}^a$ Eurasian International Center for Theoretical Physics and Department of General
Theoretical Physics, Eurasian National University, Astana 010008, Kazakhstan \\ }
 \vspace*{0.5cm}
{\it ${}^b$ Department of Science, Campus of Bijar, University of  Kurdistan  \\
Bijar, IRAN. \\ }
 \vspace*{0.5cm}
%{\it ${}^c$ PNU }

\vspace*{2cm}
%%\maketitle
\end{center}

\begin{abstract}
We present the  analytic Lifshitz solutions for a scalar field model  minimally coupled with the abelian gauge field in $N$ dimensions. We also consider the presence of cosmological constant $\Lambda$. The Lifshitz parameter $z$ appearing in the solution plays the role of the Lorentz breaking parameter of the model. We investigate the thermodynamical properties of the solutions and discuss the energy issue. Furthermore, we study the hairy black hole solutions in which the abelian gauge field breaks the symmetry near to the horizon. In the holographic picture, it is equivalent to a second order phase transition. Explicitly we show that there exists a critical temperature which is a function of the Lifshitz parameter $z$. The system below the critical temperature becomes superconductor, but the critical exponent of the model remains the same of the usual holographic superconductors without the higher order gravitational corrections, in agreement with Ginzburg-Landau theories.

\end{abstract}

\end{titlepage}

%%%%%%%%%%%%%%%%%%%
%%%%%%%%%%%%%%%%%%%%%%%%%%%%%%%%%%%%%%%%%%%%%%%%%%%%%%%%%%%%%%%%%%%%%%%%%%%%%%
%%%%%%%%%%%%%%%%%%%%%%%%%%%%%%%%%%%%%%%%%%%%%%%%%%%%%%%%%%%%%%%%%%%%%%%%%%%%%%
\section{Introduction}
The anti-de Sitter/conformal field theory (AdS/CFT) correspondence \cite{1} has many useful applications in condensed matter
physics, especially for studying scale-invariant strongly-coupled systems, for example, and low temperature systems near
quantum criticality \cite{2}. In brief, this conjecture states that there is a direct duality between any classical solution of the gravitational bulk action in $d$ dimensional asymptotically $AdS_d$ spacetime to the quantum objects on the boundary. These quantum objects are described as well by a conformal field theory. At the first, it seems this following deep conjecture supported by the relativistic essence of the bulk action. By the classical solutions in the bulk we mean those locally Lorentz invariant theories of the gravity (or modified gravity but with more precise view) not with the semi quantum corrections. So the bulk must be completely classical from this point of view. Also, the quantum object who lives on the boundary must be unitary and real operator to support a qualified quantum theory without any serious problem. Different kinds of the dualities can be addressed. One possibility is the correspondence between the gravity solutions and the fluid mechanics equations. By some techniques which are inspired directly from the holographic picture \cite{fluid} it is possible to find some more information about the exact solutions of the Navier-Stokes equations in some strictly imitated cases. Also the scale invariant condensation phenomena in the condensed matter provides another landscape for this conjecture. There are many attempts to relate the
condensed matter problems to their gravitational duals. Since the high-$T_c$ superconductors are
shown to be in the strong coupling regime, the BCS theory fails and one expects
that the holographic method could give some insights into the pairing mechanism
in the high-$T_c$ superconductors. From the ($d$ dimensional) field
theory point of view, superconductivity is characterized by condensation of a generally
composite charged operator $\hat{O}$ in low temperatures $T < T_c$. In the gravitational dual
($d+1$ dimensional) description of the system, the transition to the super conductivity is
observed as a classical instability of a black hole in an anti-de Sitter (AdS) space against
 perturbations by a charged scalar field $\psi$. The AdS/CFT correspondence relates the quantum dynamics of the boundary operator
$\hat{O}$
to a simple classical dynamics of the bulk scalar field $\psi$ \cite{3,4}. Various holographic superconductors have been
studied in Einstein theory \cite{5,6} or extended versions as Gauss-Bonnet (GB) \cite{7,8}, Weyl corrected ones \cite{mpla,epl,ijmpa,ijmpa2013}, with magnetic field in the bulk action \cite{epl2,wen,scripta,alb1,alb2,alb3,alb4}, with non linear Maxwell's fields \cite{gravity}, as a  toy model of two  dimensional superconductors using $AdS_3/cft_2$ \cite{ijtp}, in modified gravity scenario \cite{FR} and even in the non relativistic model of gravity as, for example, in Horava-Lifshitz theory \cite{9,10}. The analytical methods have been used in the description of the phase transition phenomena in the superconductors of type II \cite{dibakar}.
In recent years, holographic method have been used to study non-relativistic system \cite{11}. In the framework of condensed mater theory, different systems show a dynamical scaling near fixed points\footnote{In this paper we use a different notation with respect to the one usually adopted  in parameterizing the Lifshitz solution. In the seminal paper \cite{PRD2012}, $(z+2)/2\rightarrow z$. The reason of our choice is a more suitable form of Lifshitz solution for our derivation.}:
\begin{equation}
t\rightarrow \lambda^{\frac{z+2}{2}}t, \hspace{0.5 cm} x_i\rightarrow \lambda x_i, \hspace{0.3 cm} z\neq 0.
\end{equation}
As a consequence, instead obeying the conformal scale invariance $t\rightarrow \lambda t,  x_i\rightarrow \lambda x_i$ , the temporal and the spatial coordinates scale anisotropically. The Lifshitz topological black holes and charged Lifshitz black holes have previously been discussed in Refs. \cite{mann1,mann2}. Also the Einstein-Maxwelldilaton
system has been used to construct Lifshitz spacetime\cite{Goldstein}.  \\
In the present paper we would like to study holographic superconductors in a new background. This set up of the s-wave holographic superconductors uses the AdS-Lifshitz  black hole as the gravitational bulk metric in the probe limit. In the Section 2, we introduce a scalar field model non minimally coupled
with the abelian gauge field in the presence of cosmological
constant in $N$ dimensions and we derive static, (pseudo-)spherically symmetric (SSS) solutions
with various topologies. In particular, we will be interested in the AdS-Lifshitz  black hole (AdS-BH)  solutions.  
In the Section 3, we study the thermodynamical properties of the solutions and obtain the quasi-local generalized Misner-Sharp mass as a Killing conserved charge. We also verify the validity of the Gibbs equation
by using the 
%obtain the Wald entropy, Killing-Hawking temperature and 
Kodama-Hayward temperature.
In Section 4 we study the hairy black hole solutions in which near the horizon the abelian gauge field breaks the symmetry and in Section 5 we explore the scalar condensation in our Lifshitz black hole solutions  by analytical
approaches. The matching solutions and the critical temperature will be
found. Finally, conclusions are given in the last Section.

%%%%%%%%%%%%%%%%%%%%%%%%%%%%%%%%%%%%%%%%%%%%%%%%%%%%%%%%%%%%%%%%%
%%%%%%%%%%%%%%%%%%%%%%%%%%%%%%%%%%%%%%%%%%%%%%%%%%%%%%%%%%%%%%%%%

\section{Bulk asymptotic $ AdS_N$-Lifshitz solution }
%%%%%%%%%%%%%%%%%%%%%%%%%%%%%%%%%%%%%%%%%%%%%%%%%%%%%%%%%%%%%%%%%
%%%%%%%%%%%%%%%%%%%%%%%%%%%%%%%%%%%%%%%%%%%%%%%%%%%%%%%%%%%%%%%%%
We will consider the $N$-dimensional action of the following model where the scalar field $\phi$ is non minimally coupled with electromagnetic potential,
\begin{equation}
I=\int_{\mathcal{M}} d^{N}x\sqrt{-g}\left[\frac{(R-2\Lambda)}{2\kappa^2}-\frac{1}{2}
\partial^\mu\phi\partial_\mu\phi
+V(\phi)
-\xi\mathrm{e}^{\lambda\phi}(F^{\mu\nu}F_{\mu\nu})\right]\,.\label{action}
\end{equation}
Here, $g$ is the determinant of metric tensor, $g_{\mu\nu}$, $\mathcal{M}$
is the space-time manifold, $\Lambda$ is a "non effective" cosmological constant, namely $\Lambda=-(N-1)(N-2)/(2L^2)$, $L$ being a length size, and $F_{\mu\nu}$ is the electromagnetic field strength coupled with scalar field $\phi$ as $\xi\mathrm{Exp}[\lambda\phi](F^{\mu\nu}F_{\mu\nu})$, $\xi$ and $\lambda$ being generic constants (for example, in the four dimensional Einstein- Maxwell action one has $\xi=1/4$ and $\lambda=0$).
The field is also subjected to a potential $V(\phi)$.
Here, we use units of $k_{B}=c=\hbar=1$ and denote the gravitational constant
$\kappa^2=8\pi G_N^{N}\equiv8\pi(1/M_{Pl}^2)^N$ with the Planck mass of
$M_{PL}=G^{-1/2}_N=1.2\times 10^{19}\text{GeV}$.

We look for static, (pseudo-)spherically symmetric (SSS) solutions
with various topologies, and write the metric element as
\begin{equation}
ds^2=-e^{2\alpha(r)}B(r)dt^2+\frac{d r^2}{B(r)}+r^2d\sigma_{N-2,k}^2
\,,\label{metric}
\end{equation}
where $\alpha(r)$ and $B(r)$ are function of $r$ only and $d \sigma_{N-2,k}^2$ represents the metric of a topological $(N-2)$-dimensional surface parametrized by $k= 0, \pm 1$, such that the manifold will be either a sphere $S_{N-2}$ (for $k=1$), a torus $T_{N-2}$ (for $k=0$) or a compact hyperbolic manifold $Y_{N-2}$ (for $k=-1$). In particular, we will be interested in the Lifshitz solutions, where $\alpha(r)\propto\log r^{z/2}$, being $z$ the redshift parameter.
In this case, for power counting renormalizability in $N$ dimension, if we assume that the interaction potential can be expanded as 
\begin{equation}
V(\phi)=\Sigma_{i=0}^{K}\,g_i \phi^i\,, \nonumber
\end{equation}
where $\phi^i$ are the polynomial terms of the series and $g_i$ suitable coefficients, by the dimensional engineering we get\cite{ren}
\be
[g_i]=[m]^{\frac{[N+z-1-m(N-1-z)/2}{z}}\nonumber\,.
\ee
Such kind of theory is renormalizable if the couplings have non-negative momentum (here mass) dimension $[m]$, so that we have two possibilities, namely
\begin{eqnarray}
K=\frac{2(N-1+z)}{N-1-z},\ \ z<N-1\,,\nonumber\\
K=\infty,\ \ z\geq N-1.\nonumber
\end{eqnarray}
The above constraints are valid for any scalar theory  under the Lifshitz scaling of the coordinates.

We propose the potential in the following form,
\begin{equation}
V(\phi)=V_0\mathrm{e}^{\gamma\phi}\label{V}\,,
\end{equation}
where $V_0$ and $\gamma$ are generic parameters.
%This choice is motivated by inflationary scenarios.
Now, by comparing the exponential potential (\ref{V}) with the polynomial form, we observe that in fact $K=\infty$, so that the theory (\ref{action}) is renormalizable for $z\geq N-1$.
%This is the renormalizability condition for (\ref{action}).

With the metric Ansatz (\ref{metric}), the scalar curvature reads
\begin{eqnarray}
R  =
-3B'
\alpha'-2\,B
\alpha'^{2}-
B''-2\,B\alpha''
-(N-2)\left[
\frac{2}{r}B'
+2\,\frac{B\alpha'}{r}
-\frac{(N-3)}{r^2}(k-B)
\right]
\,,\label{R}
\end{eqnarray}
where the prime index denotes the derivative with respect to $r$.
Where is not necessary, the argument of the functions $\alpha(r)$ and $B(r)$ will be dropped.

%The field equations for Maxwell field are derived from the action (\ref{action}) and read
%\begin{equation}
%\nabla_k F^{kj}=0\,,\label{Max}
%\end{equation}
%where $\nabla_k$ is the covariant derivative operator associated with the metric.
Moreover, due to the $SO(N-2)$ symmetry and also demanding the parity symmetry, it is easy to see that the only non vanishing
components of the electromagnetic field in $N$-dimension are
\begin{equation}
F_{01}=
-\frac{d A_0}{d r}\,,
\quad
F^{01}=g^{00}g^{11}F_{01}=\text{e}^{-2\alpha}\frac{d A_0}{d r}
\,,\nonumber
\end{equation}
\begin{equation}
F_{10}=
\frac{d A_0}{d r}\,,
\quad
F^{10}=g^{11}g^{00}F_{10}=-\text{e}^{-2\alpha}\frac{d A_0}{d r}
\,,\label{relations}
\end{equation}
$A_0$ being the electric potential. 
Now, in order to find the EOMs we can use the reduced action~\cite{Deser}.
By assuming $\phi=\phi(r)$ and
by plugging the above expressions into the action (\ref{action}), making a partial integration, we finally get the following effective Lagrangian,
\begin{eqnarray}
\mathcal{L_{\mathrm{eff}}}&=&\mathrm{e}^{\alpha}r^{N-2}\left[\frac{(N-2)(N-3)(k-B)}{r^2}-
\frac{(N-2)B'}{r}-2\Lambda
-\frac{B\tilde\phi'^2}{2}+\tilde V_0\mathrm{e}^{\gamma\phi}+\right.\nonumber\\
&&\left.
2(2\kappa^2)\xi\mathrm{e}^{\lambda\phi}\text{e}^{-2\alpha}\left(\frac{d A_0}{d r}\right)^2
\right]\,,
\end{eqnarray}
where, for simplicity, we have putted $\tilde\phi=2\kappa^2\phi$ and $\tilde V_0=2\kappa^2V_0$.

The field equation for electromagnetic field coupled with the scalar field $\phi$ are derived from the action (\ref{action}) and read
\begin{equation}
\frac{d}{d r}\left(r^{N-2}\text{e}^{\lambda\phi}\text{e}^{-\alpha}\frac{d A_0}{d r}\right)=0\,,\nonumber
\end{equation}
namely
\begin{equation}
\frac{d A_0}{d r}=\frac{\text{e}^{\alpha}\text{e}^{-\lambda\phi}Q}{r^{N-2}}\,,\label{QQQ}
\end{equation}
$Q$ being an integration constant of the electromagnetic field. The identification of $Q$ with the classical electric charge (eventually multiplied to some suitable dimensional parameter) is recovered in the flat limit $\alpha=0$ and in the absence of the coupling with the field ($\lambda=0$) as a consequence of the Gauss theorem and of the vanishing of electric potential at large distances. 

By using the Euler-Lagrangian equations
$$
\frac{d}{dr}\Big(\frac{\partial\mathcal{L}}{\partial Z'_A}\Big)=\frac{\partial\mathcal{L}}{\partial Z_A}\,,
$$
where, in our case, $Z_A=\{\alpha(r),B(r)\}$, we also obtain the following equations of motion (EOMs):
\begin{equation}
B'r-(N-3)(k-B)-\frac{r^2}{(N-2)}\left[-2\Lambda-\frac{\tilde\phi'^2B}{2}+\tilde V_0\mathrm{e}^{\gamma\phi}
-\frac{2\xi\mathrm{e^{-\lambda\phi}}\tilde Q^2}{r^{2(N-2)}}\right]=0\,,\label{1}
\end{equation}
\begin{equation}
\alpha'-\frac{\tilde\phi'^2r}{2(N-2)}=0\,.\label{2}
\end{equation}
Here, after the derivation, we have used Eq.~(\ref{QQQ}) in (\ref{1}) and we have putted $\tilde Q^2=2\kappa^2 Q^2$.

%This equations correspond to the $(t,t)$ and $(r,r)$ independent components of the usual %tensorial form of field equations,
%\begin{equation}
%R_{\mu\nu}-\frac{1}{2}g_{\mu\nu}R+\Lambda g_{\mu\nu}=\kappa^2(T_{\mu\nu}^{\phi}+
%\xi\mathrm{e}^{\lambda\phi}(4F_{\nu\sigma}\partial_\mu A^{\sigma}-%g_{\mu\nu}F^{\alpha\beta}F_{\alpha\beta})
%)\,,
%\label{Field}
%\end{equation}
%where $R_{\mu\nu}$ is the $N$-dimensional Ricci tensor, $T_{\mu\nu}^{\phi}$ is the stress %energy tensor of the field $\phi$ and $A_{\mu}$ represents the electromagnetic potential such %that $F_{\mu\nu}=\nabla_\mu A_{\nu}-\nabla_\nu A_{\mu}$.

Finally, the equation for $\phi(r)$ reads
\begin{equation}
\tilde\phi''B+\left[\alpha'B+B'+\frac{N-2}{r}B\right]\tilde\phi'+\left[\gamma\tilde V_0\mathrm{e}^{\gamma\phi}+
\frac{2\xi\lambda\mathrm{e}^{-\lambda\phi}\tilde Q^2}{r^{2(N-2)}}\right]=0\,,\label{3}
\end{equation}
where we have used (\ref{QQQ}) again after the derivation.

As we stated above, we are interested in the following solutions,
\begin{equation}
\alpha(r)=\log[(r/r_0)^{z/2}]\,,\label{alpha}
\end{equation}
which correspond to the important class of Lifshitz solutions parameterized by red shift $z$ parameter. Here, $r_0$ is a dimensional constant. From Eq. (\ref{2}) we get
\begin{equation}
\tilde\phi(r)=\sqrt{z(N-2)}\log[r/\tilde r_0]\,,\label{phi}
\end{equation}
so that for renormalizable theory ($z>N-1$) we deal with real fields, since in general we take $N>3$. Here, $\tilde r_0$ is a new scale constant, which is in principle different from $r_0$ introduced above, and $\tilde\phi(r)$ has been taken with positive sign (an other possible solution is given by $\tilde\phi(r)=-\sqrt{z(N-2)}\log[r/\tilde r_0]$).
By using this result, we can solve Eq. (\ref{1}) as
\begin{eqnarray}
B(r)&=&\frac{2k(N-3)}{z+2N-6}+C\,r^{3-N+z/2}-\frac{2r^{A}\tilde V_0}{(N-2)(6-2A-2N-z)}\nonumber\\
&&+\frac{4\xi\tilde Q^2 r^{B}}{(N-2)(6-2B-2N-z)}-\frac{4\Lambda r^2}{(N-2)(2N-2+z)}\,.\label{BB}
\end{eqnarray}
In this equation, $C$ is a free integration constant of the solution and
\begin{eqnarray}
A&=&\gamma\sqrt{z(N-2)}+2\,,\nonumber\\
B&=&-\lambda\sqrt{z(N-2)}+6-2N\,.
\end{eqnarray}
For simplicity, in the above expression we have also redefined $r_0^{-\gamma\sqrt{z(N-2)}}\tilde V\rightarrow\tilde V$
and $\tilde r_0^{\lambda\sqrt{z(N-2)}}\tilde Q^2\rightarrow\tilde Q^2$.
Note that
if we turn off the scalar field potential and then take $\phi=0$, we recover the Reissner-Norstrom solution with cosmological constant for $z=0$,
\begin{eqnarray}
d s^2&=& -B(r)dt^2+\frac{d r^2}{B(r)}+r^2d\sigma_{(N-2),k}^2\,,\nonumber\\
B(r)&=&k+C r^{3-N}-\frac{2\Lambda r^2}{(N-2)(N-1)}+\frac{2\xi\tilde Q^2 r^{6-2N}}{(N-2)(N-3)}\,.
\label{vacuum}
\end{eqnarray}
On the other hand, in the presence of the scalar field,
the solution (\ref{BB}) is acceptable only if the Klein Gordon equation
 (\ref{3}) for $\phi$ is also satisfied. In this case, the generic form of the metric is given by
\begin{equation}
ds^2=-\left(\frac{r}{r_0}\right)^{z}B(r)dt^2+\frac{d r^2}{B(r)}+r^2d\sigma_{N-2,k}^2\,.\label{exsolution}
\end{equation}
Let us see for two different cases.
\begin{itemize}
\item Absence of cosmological constant.
If $\Lambda=0$, one possible solution of Eq. (\ref{3}) is found by choosing
\begin{eqnarray}
&&\gamma=-\frac{2}{\sqrt{z(N-2)}}\,,\nonumber\\
&&\lambda=-\frac{2(N-3)}{\sqrt{z(N-2)}}\,,\nonumber\\
&&\tilde V_0=\frac{k(N-3)(N-2)z-2\tilde Q^2(2N-6+z)\xi}{(2-z)}\,.
\end{eqnarray}
The solution becomes
\begin{eqnarray}
B(r)=-\frac{4[k(N-3)-2\tilde Q^2\xi]}{(z-2)(2N-6+z)}
+C r^{3-N-z/2}\,.\label{lambda0}
\end{eqnarray}
One remark is in order. In this and in the next case, the solution is not unique and depends on the choice of the parameters which must satisfy Eq. (\ref{3}). As a consequence,
we added some (suitable) additional constrains on the parameters $\gamma$ and $\lambda$. Thus, the constrain on $\tilde V_0$ follows from Eq. (\ref{3}).
\item Cosmological constant $\Lambda\neq 0$. One simple solution is given by
\begin{eqnarray}
&&\lambda=-\frac{2(N-2)}{\sqrt{z(N-2)}}\,,\nonumber\\
&&\gamma=-\frac{2}{\sqrt{z(N-2)}}\,,\nonumber\\
&&z=-\frac{2\left(N-2\right)\tilde Q^2\xi}{\Lambda+\tilde Q^2\xi}\,,\nonumber\\
&&\tilde V_0=-\frac{k(N-3)(N-2)^2\tilde Q^2\xi}{(N-1)\tilde Q^2\xi+\Lambda}\,.
\end{eqnarray}
The solution reads
\begin{eqnarray}
B(r)&=&\frac{2}{(2N-6+z)}\left[k(N-3)+\frac{\tilde V_0}{(N-2)}\right]+C r^{3-N-z/2}\nonumber\\
&&-\frac{4r^2(\Lambda+\tilde Q^2\xi)}{(N-2)(2N-2+z)}\,,\label{lambda}
\end{eqnarray}
which asymptotically is a de Sitter/Anti de Sitter (dS/AdS) solution in the case of $z=0$.
Note that if $\Lambda=0$, we get $z=-2(N-2)$, and, for $N>2$, the solution is not accettable, being $\phi$ in Eq. (\ref{phi}) imaginary, and the theory becomes non renormalizable. In principle, for any dimension $N$ and for any choice of $z$ we can obtain the corresponding Lifshitz solution by setting the values of $\gamma$, $\lambda$, $Q$ and $V_0$ appearing in the field lagrangian.

The solution (\ref{lambda}) can be asymptotically Lifshitz AdS and will furnish our bachground in studying holographic superconductors. For a planar horizon $k=0$, our solution turns out to be the one of Ref. \cite{cucu}, while the solutions for $k=-1,1$ are novel.

\end{itemize}

%%%%%%%%%%%%%%%%%%%%%%%%%%%
\section{Black hole solutions and thermodynamics}
%%%%%%%%%%%%%%%%%%%%%%%%%5

The solutions derived in the previous Section may describe charged black holes (BHs) in $N$-dimensional manifolds in the presence of scalar field non minimally coupled with electrodynamic potential. We recall that event horizon exists as soon as there exists a
positive solution $r_{+}$ of
\begin{equation}
B(r_{+})=0\,,\quad\quad B'(r_{+})\gneq 0\,.\label{BHconditions}
\end{equation}
We require $B(r_{+})\neq $ to avoid the extremal BHs.

In the case of solution (\ref{lambda0}) one has
\begin{equation}
r_{+}=\left[\frac{4k(N-3)-8\tilde Q^2\xi}{C(z-2)(2N-6+z)}
\right]^{\frac{1}{3-N-z/2}}\,.
\end{equation}
Since $z>0$, we must require $k/C>2\tilde Q^2\xi(z-2)/(N-3)$ when $N>3$. However, since in general $\xi>0$, in order to have $B'(r_{+})>0$, we see that only in the topological case $k=-1$ we obtain a BH solution.

Concerning the solutions (\ref{vacuum}) and (\ref{lambda}), it is always possible to describe topological BHs by making an appropriate choice of the parameters. For example, in the case of solution (\ref{lambda}) with
\begin{equation*}
\frac{2}{(2N-6+z)}\left(k(N-3)+\frac{\tilde V_0}{(N-2)}\right)>0\,,\quad\frac{4(\Lambda+\tilde Q^2\xi)}{(N-2)(2N-2+z)}<0\,,
\end{equation*}
 the equation $B(r)=0$ has two roots, namely $r_\pm$, the first one corresponding to the event horizon of the black hole ($B'(r_{+})>0$) and the second one to the (Anti-de Sitter) horizon of the cosmological background where the black hole is  immersed ($B'(r_{-})<0$).

In the following, we will assume to deal with solutions whose parameters satisfy the  conditions (\ref{BHconditions}) for some value of $r=r_{+}$.

Let us study some physical propriety of these black holes.
Since the field equations of the theory are second order differential equations,
we can easily derive a conserved current whose charge may be identified with the mass of the BHs. We will follow the approach proposed by Wheeler for Lovelock theories~\cite{lovelock} in Ref.~\cite{wheeler}.
For simplicity, we denote with $\mathcal G_{\mu\nu}$ the Einstein tensor plus the contribute of cosmological constant, namely
\begin{eqnarray}
\mathcal G_{\mu\nu}=R_{\mu\nu}-\frac{1}{2}g_{\mu\nu}R+\Lambda g_{\mu\nu}\,.\label{tensor}
\end{eqnarray}
By means of time-like Killing vector field $K^{\nu}=(1,\vec{0})$ in $N$-dimension,
%$K_{\mu}$, $\nabla_\mu K_{\nu}+\nabla_\nu K_\mu=0$
in the case of static metric (\ref{metric}),
one can construct the conserved current
\begin{equation}
J_\mu:=\mathcal G_{\mu\nu}K^\nu\,,
\end{equation}
such that
\begin{equation}
\nabla_\nu J^\nu=0\,,
\end{equation}
being $\mathcal G_{\mu\nu}$ a conserved quantity.
A direct evaluation of $J_{0}$ via $(\ref{tensor})$ leads to
\begin{eqnarray}
 J_0:= \mathcal G_{00}K^0=(\mathrm{e}^{2\alpha(r)}B(r))\frac{(N-2)}{2r^{N-2}}\frac{d[r^{N-1}W(r)]}{dr}\,,
\end{eqnarray}
where
\phantom{line}
\begin{eqnarray}
W(r)=[k-B(r)]r^{-2}
-\frac{2\Lambda}{(N-2)(N-1)}\,.\label{W(r)}
\end{eqnarray}
The current $J_\mu$
gives rise to a Killing conserved charge.
This corresponds to the quasi-local generalized Misner-Sharp \cite{MS} mass which reads\\
\phantom{line}
\begin{equation}
E_{MS}(r)\equiv-\frac{1}{\kappa^2}\int_\Sigma  J^\mu d \Sigma_\mu
=\frac{(N-2)V_{N-2,k}}{2\kappa^2}\int_0^r d\rho \frac{d(\rho^{N-1}W)}{d \rho}
=\frac{(N-2)V_{N-2,k}}{2\kappa^2}r^{N-1}W(r)\,,
\label{EMS}
\end{equation}
\phantom{line}\\
where $\Sigma$ is a spatial volume at fixed time,  $d\Sigma_\mu=(d\Sigma, \vec{0})$, and
$V_{N-2,k}$ is the $N-2$ dimensional volume depending on the topology. For example, in the case of the sphere with $k=1$, one has $V_{N-2,1}=2\pi^{(N-1)/2}/\Gamma((N-1)/2)$, with $\Gamma(z)$ the Euler-Gamma function.

In particular, on shell, that is at the horizon $r=r_{+}$  such that $B(r_{+})=0$, the quasi local energy is identified with the black hole energy $E$ which reads
\begin{equation}
E:=E_{MS}(r_{+})=\frac{(N-2)V_{N-2,k}}{2\kappa^2}\left(k r_{+}^{N-3}-\frac{2\Lambda}{(N-2)(N-1)}r_{+}^{N-1}\right)\,.
\label{E}
\end{equation}
For example, in the vacuum case of
Eq. (\ref{vacuum}) with $\tilde Q=0$
one has
\begin{eqnarray}
r^{N-1}W(r)=-C\,,
\label{C}
\end{eqnarray}
and the black hole energy reads
\begin{equation}
E=-\frac{(N-2)V_{N-2,k}}{2\kappa^2}C \,,
\label{E2}
\end{equation}
so that the constant of integration is related with the mass of the BH.

We note that expression ($\ref{E}$) correctly returns the Misner-Sharp mass for asymptotically flat solutions ($\Lambda=0$) in vacuum or in the presence of matter. In particular, for $N=4$ and $k=1$, by explicitly writing the Newton Constant, we get the familiar result $E=r_{+}/(2 G_{\mathrm{N}})$, which corresponds, in the vacuum case, to $E=-C/(2G_{\mathrm{N}})$.
%On the other hand, in order to have BH solutions with positive energy in the topological cases %$k=0,-1$, we
%need a cosmological constant, such that $B'(r_{+})>0$.

Now, let us show that the Gibbs equation $TdS=dE-pdV$ holds true for the black holes described by the model (\ref{action}), with the Killing energy $E$
obtained below,  and the pressure $p$ given by electromagnetic and scalar fields. $T$ and $S$ are the temperature and the entropy of the black hole, and $V$ is the volume enclosed by the horizon in $N-1$ dimensional space.

For Lovelock gravity the validity of the First Law of black hole thermodynamics
has been investigated
in several places~\cite{meyer,pad2,maeda}.
For our static non vacuum case we present a simple derivation from the first EOM (\ref{1}) evaluated on the horizon $r=r_{+}$,
\begin{eqnarray}
&&\frac{(N-2)V_{N-2,k}B'(r_{+})r_{+}^{N-3}}{2\kappa^2}-\frac{(N-2)V_{N-2,k}}{2\kappa^2}\left(k r_{+}^{N-3}-\frac{2\Lambda}{(N-2)(N-1)}r_{+}^{N-1}\right)\nonumber\\
&&-\frac{V_{N-2,k}r^{N-2}}{2\kappa^2}
\left(\tilde V_0 \mathrm{e}^{\gamma\phi}-\frac{2\xi\mathrm{e}^{-\lambda\phi}\tilde Q^2}{r_+^{2(N-2)}}\right)=0\,.\label{prima}
\end{eqnarray}
Here, we have used the fact that $B(r_{+})=0$.

All thermodynamical quantities associated with black holes solutions
can be computed by standard methods. The entropy can be calculated by the Wald method~\cite{Wald,Visser:1993nu,FaraoniEntropy} and reads
\begin{equation}
S_W=\frac{2\pi V_{N-2,k}}{\kappa^2}r_{+}^{N-2}\,.
\label{entropy}
\end{equation}
Furthermore, for the static metric (\ref{metric}) it is possible to find a characteristic temperature related to the event horizon. A natural choice is to take the so called Killing/Hawking temperature \cite{HT}
\begin{equation}
T_K:=\frac{\kappa_K}{2\pi}=\frac{\mathrm{e}^{\alpha(r_{+})}}{4\pi}B'(r_{+})\,,\label{K}
\end{equation}
whose validity may be justified
making use of derivations of Hawking radiation~\cite{Visser2}
or by eliminating the conical singularity in the corresponding Euclidean metric \cite{Roberto}
or making use of  the tunneling method~\cite{PW,Nadalini}.
%This is a well known result, and it can be justified in several ways,
%for example  making use of standard derivations of Hawking radiation~\cite{Visser2},
%or by eliminating the conical singularity in the corresponding Euclidean metric,
%or making use of  the tunneling method,
%recently introduced in Refs.~\cite{PW,Nadalini}, and discussed in details in several papers.
In the above expression, $\kappa_K$ denotes the Killing surface gravity, namely $\kappa_K=\mathrm{e}^{\alpha(r_{+})}B'(r_{+})/2$, 
derived from the relation 
$K^{\mu}\nabla_\mu K^\nu=\kappa_K K^\nu$, where 
$K^{\nu}=(1,\vec 0)$ is the time-like Killing vector field.

However, we would like to remind that in the spherical symmetric, dynamical case, the real geometric object which generalizes the Killing vector field is
the Kodama field~\cite{Kodama} 
with a related conserved current and a related Kodama surface gravity $\kappa_H$. In such a case, a natural definition of the temperature for dynamical black holes reads as $T_H=\kappa_H/2\pi$, where $T_H$ is the Kodama/Hayward temperature, in analogy with the static case. This temperature permits to find the Gibbs relation in the dynamical case (see the seminal work of Hayward in Ref. \cite{H} and the Appndix A), when the Killing surface gravity cannot be defined being the time-like Killing vector field absent. The interesting point is that in the static case the Kodama vector field still exists but does not coincide with the time-like Killing vector field
and differs from it as $\mathcal K^\nu=\mathrm{e}^{-\alpha(r)} K^\nu$, namely
\begin{equation}
\mathcal K^\mu=\left(\mathrm{e}^{-\alpha(r)}, \vec 0\right)\,,
\end{equation}
such that the Kodama/Hayward surface gravity reads 
\begin{equation}
\kappa_H=\frac{B'(r_+)}{2}\,.
\end{equation}
As a consequence, one finds
\begin{equation}
T_H:=\frac{\kappa_H}{2\pi}=\frac{1}{4\pi}B'(r_{+})\,,\label{HH}
\end{equation}
namely $T_H=\mathrm{e}^{-\alpha(r)} T_K$ and in principle definitions (\ref{K}) and (\ref{H}) are  different. In vacuum case where $\alpha(r)=0$, the two temperatures coincide, but for "dirty" BHs (i.e., in the presence of matter) as the ones we are considering, $T_K\neq T_H$.
This is related with the fact that the Killing vector cannot be defined unambiguously when the space-time is not asymptotically flat.
We stress that all derivations of Hawking radiation
lead to a
semi-classical expression for the black hole radiation rate $\Gamma$,
\begin{equation}
\Gamma\equiv \mathrm{e}^{-\frac{\Delta E_K}{T_K}}\,,
\label{ratekill}
\end{equation}
in terms of the change $\Delta E_K$ of the Killing energy $E_K$,
but if one uses the Kodama energy $E_H$ for the emitted particle, one has
\begin{equation}
\Gamma\equiv \mathrm{e}^{-\frac{\Delta E_H}{T_H}}\,.
\label{ratekoda}
\end{equation}
This fact derives by the relationship $\Delta E_H=\mathrm{e}^{-\alpha(r)}\Delta E_K$.
From the Eqs. (\ref{ratekill})-(\ref{ratekoda}), one arrives at the identity
\begin{equation}
\label{kk}
\frac{\Delta E_H}{T_H}=\frac{\Delta E_K}{T_K}\,,
\end{equation}
so that the tunneling probability is invariant under different choices of the temperature.

In Ref. \cite{Cognola} an attempt to identify the mass of static BHs in modified theories of gravity as the integration constant which appears in the vacuum solutions has been done.
This result has been derived by the EOMs and seems in favor of the Killing temperature with respect to the Kodama-Hayward one, but here, for our non vacuum solutions, Eq. (\ref{prima}) suggests the use of the Kodama temperature (\ref{H}), 
in the attempt to recover the Gibbs relation. 
In fact, by making use of the BH entropy (\ref{entropy}), one can rewrite Eq. (\ref{prima}) as
\begin{equation}
T_H d S_W=d E+pdV\,,
\end{equation}
where $E$ is the BH energy (\ref{E}), $V$ is the volume enclosed by the horizon in $N-1$-dimensional space, $V=V_{N-2,k} r^{N-1}/(N-1)$, and $p=p_\phi+p_{EM}$ is the working term given by the radial scalar field pressure ($p_\phi$) and the radial pressure of electromagnetic field coupled with scalar field ($p_{EM}$) on the horizon
\footnote{The stress energy tensor of our model is given by $T_{\mu\mu}\equiv\mathcal{L}g_{\mu\nu}-2\partial\mathcal L/\partial g_{\mu\nu}=\partial_\mu\phi\partial_\nu\phi/2+Vg_{\mu\nu}+4\xi\text{e}^{\lambda\phi}
\left[F_\mu^\alpha F_{\nu\alpha}-g_{\mu\nu}F_{\beta\gamma} F^{\beta\gamma}/4\right]$. Given our SSS metric and the static electromagnetic field, the radial pressure is derived as $p\equiv -T_r^r= -B(r)\phi'(r)^2/2+V(r)-\xi\text{e}^{\lambda\phi}F^{\mu\nu}F_{\mu\nu}$. On the horizon, the dependence on $\phi'(r)$ drops down and we recover the expressions in the formula.
},
\begin{eqnarray}
p_\phi&=&-V(\phi)=-V_0\mathrm{e}^{\gamma\phi}\,,\nonumber\\
p_{EM}&=&-\xi\text{e}^{\lambda\phi} F^{\mu\nu}F_{\mu\nu}=\frac{2\xi\text{e}^{-\lambda\phi} Q^2}{r^{2(N-2)}}\,.
\end{eqnarray}
Here, we have reintroduced $V_0=\tilde V_0/(2\kappa^2)$ and $Q^2=\tilde Q^2/(2\kappa^2)$.
In this case,
the  Gibbs equation  holds true.

We prefer to use the Hayward temperature $T_H$ (\ref{HH} ) in analogy with therodynamic, by starting from the robust defininitions of the energy as the charge of a conserved current and the entropy via Wald method.\par
About the possible phase transitions , we would like also to mention that, at least when $k = 1$ and $z = 0$, which is the well-known AdS-solution with a spherical horizon, the system should exhibit a Hawking-Page phase transition, namely
a first-order phase transition between
thermal AdS-space and a Schwarzschild-AdS black hole. 
In our AdS-Lifshitz black hole
case, this transition is a  Hawking-Page like phase transition between large
Lifshitz black holes at high temperature and ‘thermal’ Lifshitz
(pure Lifshitz space with compact Euclidean time) at low temperature~\cite{Iran1}. 
This phase transition
corresponds to the confinement/deconfinement phase transition in dual theory. In Ref.~\cite{Iran2}, the AdS-
soliton solution from the planar black hole and black brane metric, by a double
Wick rotation, are investigated and it is shown that exists a critical temperature, where both solutions have the same free energy: at this point, a first 
order phase transition between the soliton and black hole occurs. This effect is the analogous
to the Hawking-Page transition, and in the dual field theory
this is a confinement/deconfinement  transition.

%%%%%%%%%%%%%%%%%%%%%%%%%%%%%%%%%%%%%%%%%%%%%%%%%%%%%%%%%%%%%%%%
\section{s-wave Holographic superconductors in probe limit}
%%%%%%%%%%%%%%%%%%%%%%%%%%%%%%%%%%%%%%%%%%%%%%%%%%%%%%%%%%%%%%%%
Our goal in the following sections are  to apply the black hole solution with the Lifshitz scaling which it has been obtained  before to study the holographic picture of  superconductor via gauge/gravity duality. In brief, by holographic superconductor we mean a condensed matter system under second order phase transition whose physical properties can be described by studying the dynamics of gauge field on the black hole background in the bulk.
It is proved that by direct applying the AdS/CFT conjecture one can interpret the asymptotic behavior of the gauge fields on the AdS boundary as the expectation values of some physical operators. The expectation values of such scalar operators are dual to the super current in the s-wave high temperature type II superconductors.
 Here, s-wave refers  to a scalar order parameter, whose expectation value breaks the $U(1)$ but
not the rotational symmetry. The gauge field can be $SU(2)$ and corresponds to the Yang-Mills fields. Such holographic models are called p-wave, because super current is a vector and the condensation happens usually for one homogeneous component of it. To apply the gravitational model to the superconductors, we will modify our gravitational model (\ref{action}) by adding a new matter field
subjected to some abelian gauge field. By starting from the backgound metric of gravitational action previously studied, we will investigate the scalar condensation of the new field and we will show that some phase transition occurs.

We easily see that
Eq. (\ref{lambda}) may describe
 an Lifshitz (dS/AdS) black hole solution for our non minimally coupled gravity model in $N$ dimensional bulk. We are interested in Anti de Sitter solutions.  If we want to relate our gravitational system with a strongly correlated system in the dual quantum theory, we need to describe the dual quantum operators via CFT. The condensed matter system dual to our classical BH solution can be addressed by holographic superconductors. In what follows, we will study the formation of the hairy BHs. The phenomenon is given by a second order phase transition and can be described by the holographic methods of the AdS/CFT.

At first, we note that the solution (\ref{lambda}) with $\Lambda\neq0$  may be asymptotically topological Lifshitz and it can be considered as the gravitational part of the holographic superconductor in the bulk. In fact, we take the gravity bulk as the charged topological black hole with a non zero, negative effective cosmological constant
\begin{equation}
\Lambda_{eff}=-\frac{12(\Lambda+\tilde Q^2\xi)}{(N-2)(2N-2+z)}<0\,.
\end{equation}
It is very interesting that the $U(1)$ reduced charge $\tilde{Q}$ is combined with cosmological constant in the solution, producing the AdS background. It is useful to rewrite the solution (\ref{lambda}) as
\begin{eqnarray}
B(r)=\frac{2}{(2N-6+z)}(k(N-3)+\frac{\tilde V_0}{(N-2)})+C r^{3-N-z/2}+\frac{r^2}{l_{eff}^2}\,,
\end{eqnarray}
where
\begin{eqnarray}
l_{eff}=\frac{1}{2}\sqrt{\frac{(N-2)(2N-2+z)}{\tilde{Q}^2\zeta+\Lambda}}\label{leff}
\end{eqnarray}
is the effective length scale.
We note that for $N=4$ and  $z=0$, solution (\ref{leff}) reduces to the usual Schwarzschild-AdS form with $l_{eff}=\sqrt{3/\Lambda}$. 
%The effective radius works properly for us. 
In fact, when we turn off the electromagnetic field and also we presrve the Lorentz invariance by $z=0$, the form of the effective length scale is the same as the one of an AdS uncharged black hole. 
We want to find the second order phase transition in the bulk theory by studying the boundary operators. In particular, we want to study the role of the Lifshitz scaling $z$ related to the critical temperature $T_c$ and the condensation of the dual operators $<\cal O_{\pm}>$ (see also Refs. \cite{jpa,jpa2, PRD2012}). As a starting point, in order to discuss the superconducting phase via holographic picture, we need a scalar field $\psi(r)$, with mass above the Breitenlohner-Freedman (BF) bound \cite{BF}, and an abelian gauge field $\mathcal A_\mu$, which is  minimally coupled with the scalar field, so that we modified the (\ref{action}) by introducing a new matter Lagrangian in the following form \cite{HHH}
\begin{eqnarray}
\mathcal{L}_m=-\frac{1}{4}\mathcal{F}^{\mu\nu}\mathcal F_{\mu\nu}-|D_\mu \psi|^2-m^2\psi^2\label{lm}\,,
\end{eqnarray}
where $D_\mu$ is the covariant derivative, $D_\mu=\partial_\mu-i q \mathcal A_\mu$, and
%$q$ the dyonic charge and $\mathcal A_\mu$ is a new electromagnetic field with 
$\mathcal F_{\mu\nu}$ is the electromagnetic field strength related to the abelian field. 
We note that in the proble limit and in the normal phase, when $\psi=0$, the electromagnetic field satisfies the bulk Maxwell's equations .
As a consequence, the total action results to be
\begin{eqnarray}
I_{total}=-\int d^{N}x \sqrt{-g}\mathcal{L}_m+\int_{\mathcal{M}} d^{N}x\sqrt{-g}\left[\frac{(R-2\Lambda)}{2\kappa^2}-\frac{1}{2}
\partial^\mu\phi\partial_\mu\phi
+V(\phi)
-\xi\mathrm{e}^{\lambda\phi}(F^{\mu\nu}F_{\mu\nu})\right]\,.\nonumber\\
\end{eqnarray}
In the above expressions, $q$ plays the role of bulk electric charge because $q$ appears in the covariant derivative exactly as a standard electric charge. The matter action (\ref{lm}) is different from the bulk action. In fact, we add here the minimally coupled Maxwell field in order to break the $U(1)$ symmetry of the abelian field near the BH horizon. Thus, we will work in the so called probe limit $q\rightarrow\infty$, ignoring the back reaction, in the normal phase $\psi=0$. In this case, the gravity sector decouples from the abelian one and the background metric can be derived from Eq. (\ref{exsolution}) and Eq (\ref{lambda}). 

In the probe limit the EOMs for $\psi(r)$ and $\mathcal{A}_\mu=\phi(r)\delta_{\mu t}$, $\delta_{\mu\nu}$ being the Kroenecker delta function and $\phi(r)$ a general function of $r$, read in the following forms
\begin{eqnarray}
D_{\mu}D^{\mu}\psi-m^2\psi=0,\\
\nabla^\mu\mathcal{F}_{\mu\nu}=i q[\psi^{*}D_{\nu}\psi-\psi D_\nu^*\psi^*]\,.
\end{eqnarray}
Here, $\mathcal{F}_{\mu\nu}=\partial_\mu \mathcal A_\nu-\partial_\nu \mathcal A_\mu$ and we take $\psi^*=\psi$, motivated by the fact that we are free to choose our gauge.  
%and $D_{\mu}=\partial_{\mu}-iq\mathcal{A}_{\mu}$ with respect to the metric %(\ref{exsolution}).
The above equations read in terms of the background metric as
\begin{eqnarray}
\phi''+\Big(\frac{N-2-\frac{z}{2}}{r}\Big)\phi'-\frac{2q^2\psi^2}{B}\phi=0\label{phi-R}\,,\\
\psi''+\Big(\frac{N-2+\frac{z}{2}}{r}+\frac{B'}{B}\Big)\psi'+\Big(-\frac{m^2}{B}+\frac{q^2r^{-z}\phi^2}{B^2}\Big)\psi=0\label{psi-R}.
\end{eqnarray}
%Here, $\phi\equiv\phi(r)$ and $\psi\equiv\psi(r)$. 
For $N=4$ and $z=0$ the model is perfectly described by numerical methods \cite{HHH}. We are interesting in the cases of $z\neq0$ and $N>3$.

We need to introduce in Eqs. (\ref{phi-R})-(\ref{psi-R}) the horizon radius $r_+$. We rewrite the solution in the following form,
\begin{eqnarray}
B(r)=\tilde{k}\left[1-\left(\frac{r}{r_+}\right)^{3-N-\frac{z}{2}}\right]+\Big(\frac{r_+}{l_{eff}}\Big)^2\left[\left(\frac{r}{r_+}\right)^2-\left(\frac{r}{r_+}\right)^{3-N-\frac{z}{2}}\right]\,,
\end{eqnarray}
where
\begin{eqnarray}
\tilde{k}=\frac{2}{2N-6+z}\Big(k(N-3)+\frac{\tilde{V}_0}{N-2}\Big)\,.
\end{eqnarray}
It is more
convenient to work in terms of the dimensionless parameter
$y(r)=r_+/r$, such that $y(r_+)=1$ and at the infinity $y(r\rightarrow+\infty)\rightarrow 0^+$.\footnote{In the literature usually the symbol is $z$, but here we kept $z$ for Lifshitz scaling and we introduced $y$ as the dimensionless radial coordinate.} In this case, the equations of motion
(\ref{phi-R}) and (\ref{psi-R}) can be expressed as:
\begin{eqnarray}
\phi''-\Big(\frac{\frac{z}{2}+N-4}{y}\Big)\phi'-\frac{2r_{+}^2\psi^2}{y^4B}\phi=0  \label{EOMz1}\,,\\
\psi''-\Big(\frac{\frac{-z}{2}+N-4}{y}-\frac{B'}{B}\Big)\psi'-\frac{r_{+}^2}{y^4}\Big(\frac{m^2}{B}-\frac{y^z\phi^2}{r_{+}^z B^2}\Big)\psi=0\,,   \label{EOMz2}
\end{eqnarray}
and the metric function reads
\begin{eqnarray}
B(y)=\tilde{k}\Big(1-y^{N-3+\frac{z}{2}}\Big)+\Big(\frac{r_+}{l_{eff}}\Big)^2\Big(y^{-2}-y^{N-3+\frac{z}{2}}\Big)\,.\label{B}
\end{eqnarray}
Now, the prime denotes the derivative with respect to $y=y(r)$. We have obtained the basic set up for the holographic superconductors.

%%%%%%%%%%%%%%%%%%%%%%%%%%%%%%%%%%%%%%%%%%%%%%%%%%%%%%%%%%%%%%%%%%%%%%%%%%%%
\section{Critical temperature and condensation values by matching method}
%%%%%%%%%%%%%%%%%%%%%%%%%%%%%%%%%%%%%%%%%%%%%%%%%%%%%%%%%%
We are going to calculate the condensate $\langle {\cal O} \rangle$
for fixed charge density.

%We consider the boundary conditions with these new variables.
Regularity at the horizon, namely $y=1$, requires
\begin{eqnarray}
\phi(1)=0\,,\hspace{1cm}\psi^\prime(1)B'(1)=r_{+}^2m^2\psi(1)\,.
\label{regularity}
\end{eqnarray}

We want to find approximate solutions around the horizon and asymptotically AdS limit, $y=1$ and $y=0$,
using Taylor's expansion, then  we want to connect these solutions in an arbitrary matching point $y_m$
between $y=1$ and $y=0$. In principle, we may do the computations by numerical methods using shooting algorithm, but in this paper we would like to keep the level of the analytical approach. At first, we calculate $B'(1)$ ($=(d B(r_+)/d y$) directly from Eq. (\ref{B}),
\begin{eqnarray}
B'(1)&=&-\tilde{k}(N-3+\frac{z}{2})+\Big(\frac{r_+}{l_{eff}}\Big)^2\left(-N+1-\frac{z}{2}\right)\label{B'1}.
\end{eqnarray}
It is easy to see that Eq. (\ref{B'1}) is related to the Kodama temperature (\ref{H}) by $B'(1)=-4\pi r_{+} T_H$. Consequently, by solving this equation for $r_+$ and using (\ref{B'1}) we have
\begin{eqnarray}
\frac{r_+}{l_{eff}}={\frac {  4\,T_H\pi \,l_{eff}+\sqrt {16\,{T_H}^{2}{\pi }^{2}{l_{eff}}^{2}+8\,z\tilde{k}+
16\,N\tilde{k}-12\tilde{k}-{z}^{2}\tilde{k}-4\,zN\tilde{k}-4\,{N}^{2}\tilde{k}} }{2\,N+z-2}}\label{r+}
\end{eqnarray}

 We will use (\ref{B'1},\ref{r+}) to construct the series solutions for our topological holographic superconductor in the probe limit. Specially, because now the radius of the horizon is a function of $(l_{eff},T_H,\tilde{k})$ , as a result we will show that the critical temperature depends on the toplogical parameter $\tilde{k}$ as well as the Lifshitz scaling parameter $z$. This is one of the most important results of our calculations which we demonstrate that in the Lifshitz backgrounds, the critical temperature depends on the topological parameter $\tilde{k}$.
%Our results are the generalizations of the works on holographic superconductors with Lorentz %invariance $z=0$ \cite{kuang} to the case of a charged topologically Lifshitz backgrounds with %scaling parameter $z\neq0$.

%%%%%%%%%%%%%%%%%%%%%%%%%%%%%%%%%%%%%%%
\subsection{Solution near the BH horizon}
%%%%%%%%%%%%%%%%%%%%%%%%%%%%%%%%%%%%%%%%%%%
We expand $\phi$ and $\psi$ near the black hole horizon at
$y=1$ as
\begin{eqnarray}
\phi(y)&=&\phi(1)-\phi^\prime(1)(1-y)+\frac{1}{2}\phi^{\prime\prime}(1)(1-y)^2
+\cdots\label{series1},\\
\psi(y)&=&\psi(1)-\psi^\prime(1)(1-y)+\frac{1}{2}\psi^{\prime\prime}(1)(1-y)^2
+\cdots\label{series2}.
\end{eqnarray}
From the boundary condition, we know that $\phi(1)=0$ and for simplicity we put $a:=-\phi^\prime(1)<0$ and
$b:=\psi(1)>0$ for the positivity of $\phi(y)$ and $\psi(y)$.

In order to discuss phase transition near the critical points, we need just to keep the second order terms in those series.

First, we compute the 2$^{\rm nd}$ order coefficient of $\phi$ by using Eq. (\ref{EOMz1})\,,
\begin{eqnarray}
\phi''(1)&=&-\phi'(1)\Big[-(\frac{z}{2}+N-4)+\frac{2r_{+}^2b^2}{B'(1)}\Big].
\end{eqnarray}
In this case Eq. (\ref{series1}) reads 
\begin{eqnarray}
\phi(y)=a\Big[(1-y)+\frac{1}{2}\Big[-(\frac{z}{2}+N-4)+\frac{2r_{+}^2b^2}{B'(1)}\Big](1-y)^2\Big]\label{seriesphi}.
\end{eqnarray}
We can calculate the 2nd derivative of $\psi$ from (\ref{EOMz2}) in the same way,
\begin{eqnarray}
\psi''(1)&=&-\frac{ba^2(r_+)^{2-z}}{(B'(1))^2}\,,
\end{eqnarray}
and  write the following series solution for scalar field $\psi$ (\ref{series2})
\begin{eqnarray}
\psi(y)&=&b\Big[1-\frac{m^2r_{+}^2}{B'(1)}(1-y)-\frac{a^2r_+^{2-z}}{2(B'(1))^2}(1-y)^2\Big]\label{seriespsi}\,.
\end{eqnarray}
In the above equations, $B'(1)$ is given by Eq. (\ref{B'1}).

%%%%%%%%%%%%%%%%%%%%%%%%%%%%%%%%%%%%%%%%%%%%%%%%%%%%%%%%%%
\subsection{Solution near the asymptotic AdS region}
%%%%%%%%%%%%%%%%%%%%%%%%%%%%%%%%%%%%%%%%%%%%%%%%%%%%%%%%
The asymptotic regime of the metric function $B(y)$ in the AdS boundary is completely independent  on the dimension of the spacetime $N$. It is trivial to recover the following asymptotic behavior of the metric
\begin{eqnarray}
B(y)\sim y^{-2}\,.
\end{eqnarray}
Here we take $z>0$ and $N>3$ to avoid the problems of  logarithmic divergence in $AdS_3/CFT_2$. 
%However, for $N=3$ there is a logarithmic divergence which must be regularized. 
Consider now the weak field behavior ($r\rightarrow+\infty$) of $\phi$ which depends on the value of dimension $N$. 
It is easy to show that this behavior completely changes at the critical dimension
$N_c=3-\frac{z}{2}$, namely
\begin{eqnarray}
\phi(y)&=&\frac{c_1}{\frac{z}{2}+N-3}y^{\frac{z}{2}+N-3}+c_2,\ \ N\neq N_c,\label{phiasym}\\
\phi(y)&=&c_1\log(y)+c_2,\ \ N=N_c.
\end{eqnarray}
We are interested in the fields with fall off behaviors near $y=0$, so that we take $N\neq N_c$. It is useful to write (\ref{phiasym}) in terms of the radial coordinate $r$,
\begin{eqnarray}
\phi(r)&=&\frac{c_1 r_+^{\frac{z}{2}+N-3}}{(\frac{z}{2}+N-3)r^{\frac{z}{2}+N-3}}+c_2.
\end{eqnarray}
By writing this solution in terms
of the dual physical quantities chemical potential $\mu$ and charge density $\rho$, we obtain
\begin{eqnarray}
\phi(y)&=&\mu-\frac{\rho}{r_{+}^{\frac{z}{2}+N-3}}y^{{\frac{z}{2}+N-3}}\label{ads1}\,,
\end{eqnarray}
where
\begin{equation}
\rho=-\frac{c_1 r_+^{\frac{z}{2}+N-3}}{(\frac{z}{2}+N-3)}\,,\quad\mu=\frac{\rho}{r_+^{\frac{z}{2}+N-3}}\,.\label{davood}
\end{equation}
The second condition derived from the fact that $\phi(r_+)=0$.
Also for the scalar field, near the AdS boundary we can write
\begin{eqnarray}
\psi(r)&=&\frac{<\cal O_{\pm}>}{r^{\Delta_{\pm}}}+...\label{ads2}\,,
\end{eqnarray}
where
$\cal O_{+}$ and $\cal O_{-}$ are the operators on the boundary and the conformal dimension is
\begin{equation}
\Delta_{\pm}=\frac{1}{2}[(N-1)\pm\sqrt{(N-1)^2+4m^2l_{eff}^2}]\,.
\end{equation}
Here, the mass square must satisfies the following relation
\begin{eqnarray}
m^2l_{eff}^2>-\frac{(N-1)^2}{4}\,.
\end{eqnarray}
A simple check shows that  under this mass bound, only  $\Delta_{+}$ has an enough rapid fall off and the scalar field behaves as
\begin{eqnarray}
\psi(r)\rightarrow   r^{-(\Delta_{+}+2)}+ r^{-\Delta_{+}}<\cal O_{+}>\,.
\end{eqnarray}
It means that the scalar field is dual to a quantum operator  on the boundary with conformal dimension $\Delta_{+}$ and we can ignore $\cal O_{-}$.
 This is not the unique possible choice. It is easy to find that if 
\begin{eqnarray}
-\frac{(N-1)^2}{4}< m^2l_{eff}^2<1-\frac{(N-1)^2}{4}\,,
\end{eqnarray}
both of the terms with conformal dimensions $\Delta_{\pm}$ fall off and we can keep they.  In conclusion, the quantization scheme is a valid procedure. In any case, the scalar field is asymptotic
to  $<\cal{O_{\pm}}>$ and these are dual to  operators with dimension $\Delta_{\pm}$. In fact,it is possible to write this quantization scheme in terms of the $z$ parameter as it has been proposed Ref.~\cite{jpa}. However, in this work we assume that  $-\frac{(N-1)^2}{4}\leq m^2l_{eff}^2$, so that we limit ourselves to the fall off with $\Delta_{+}$.

%%%%%%%%%%%%%%%%%%%%%%%%%%%%%%%%%%%%%%%%%%%%%%%%%%%%%%%%%%%%%%%%%
\subsection{Matching and phase transition}
%%%%%%%%%%%%%%%%%%%%%%%%%%%%%%%%%%%%%%%%%%%%%%%%%%%%%%%%%%%%%%%%%%
In this Section, we will connect the solutions (\ref{seriesphi}) and
(\ref{seriespsi}) with (\ref{ads1}) and (\ref{ads2}) at some completely arbitrary matching point
$y=y_m$. In order to connect those solutions smoothly, we require
the following four conditions:
\begin{eqnarray}
&&\mu-\frac{\rho}{r_{+}^{z/2+N-3}}y_{m}^{z/2+N-3}=a\Big[(1-y_m)+\frac{1}{2}\Big[-(\frac{z}{2}+N-4)+\frac{2r_{+}^2b^2}{B'(1)}\Big](1-y_m)^2\Big]\,,
 \label{c:phi}\\
&&-\frac{\rho(z/2+N-3)}{r_{+}^{z/2+N-3}}y_{m}^{z/2+N-4}=a\Big[-1-\Big[-(\frac{z}{2}+N-4)+\frac{2r_{+}^2b^2}{B'(1)}\Big](1-y_m)\Big],
 \label{c:dphi}\\
&&\frac{<\cal O_{+}>}{r_{+}^{\Delta_{+}}}y_{m}^{\Delta_{+}}=b\Big[1-\frac{m^2r_{+}^2}{B'(1)}(1-y_m)-\frac{a^2r_+^{2-z}}{2(B'(1))^2}(1-y_m)^2\Big]\,,
 \label{c:psi}\\
&&\frac{\Delta_{+}<\cal O_{+}>}{r_{+}^{\Delta_{+}}}y_{m}^{\Delta_{+}-1}=b\Big[\frac{m^2r_{+}^2}{B'(1)}+\frac{a^2r_+^{2-z}}{(B'(1))^2}(1-y_m)\Big]
 \label{c:dpsi}.
\end{eqnarray}
By combining Eqs. (\ref{c:phi})-(\ref{c:dphi}) 
we can eliminate $a b^2$ and one has
\begin{eqnarray}
&&\hspace{-1cm}\mu=\,{\frac {2\,\rho\,\Big(( \frac{z}{2}+N-3) {y_{{m}}}
^{\frac{z}{2}+N-3}-( -5+\frac{z}{2}+N ) {y_{{m}}}^{\frac{z}{2}+N-2}
\Big) {r_{+}}^{-\frac{z}{2}-N+3}+2\,( 1-y_{{m}} ) y_{{m}}a}{4y_{{
m}}}}
\label{mu},\\
&&\hspace{-1,5cm}b=\frac{\sqrt {2}}{2r_+}\,\sqrt {{\frac { \Big[\rho\,{y_
{{m}}}^{\frac{z}{2}+N-3} (\frac{z}{2}+N-3 )  -\Big(  ( \frac{z}{2}+N-4
 ) y_{{m}}-\frac{z}{2}-N+5 \Big) y_{{m}}a r_{+}^{\frac{z}{2}+N-3} \Big] {\it B'(1)}}{ar_{+}^{\frac{z}{2}+N-3}y_{{m}} (1- y_{{m}} ) }}}\,.
\label{b}
\end{eqnarray}
The above relations allude to the phase transition, namely, given $\rho$,
$\mu$ has a maximum value when we assume the non-trivial solution $b\neq 0$.
Now we can reveal the phase transition 
in our simple system. In order to evaluate the expectation value of the operator $<\cal O_{+}>$, we
eliminate the $a^2b$ term from (\ref{c:psi}) and (\ref{c:dpsi}) and obtain
\begin{eqnarray}
<\cal O_{+}>&=&-{\frac {{r_+}^{\Delta}b ( {m}^{2}{r_+}^{2}(y_{{m}}-1)+2
 ) {y_{{m}}}^{1-\Delta}}{( (\Delta-2)y_{{m}}-
\Delta ) }}
  \ . \label{relation:2}
\end{eqnarray}
For non-vanishing $b$, we can compute $<\cal O_{+}>$ from Eqs. (\ref{c:psi})-(\ref{c:dpsi}) and one gets
\begin{eqnarray}
&&a=|B'(1)|\sqrt {\frac{2\Delta^2-m^2r_{+}^2\Big(y_m+2\Delta(1-y_m)\Big)}{r_{+}^{2-z}\Big[\Delta(1-y_m)^2+2y_m(1-y_m)\Big]}}
.
\end{eqnarray}
By plugging this result in Eq. (\ref{b}) we derive
\begin{eqnarray}
&&\hspace{-1cm}b=\frac{1}{2r_+}\sqrt{\frac{\sqrt{2}B'(1)}{(1-y_m)\Sigma r_{+}^{\frac{z}{2}+N-3}}\Big[(\frac{z}{2}+N-3)\rho y_{m}^{\frac{z}{2}+N-4}+\sqrt{2}r_{+}^{\frac{z}{2}+N-3}\Sigma\Big((\frac{z}{2}+N-4)(1-y_m)-1\Big)\Big]}\,,                   \nonumber\\\\
&&\Sigma=|B'(1)|\sqrt {{\frac {{\it }\, ( {m}^{2}{r_{+}}^{2} ( \Delta-1
 ) y_{{m}}-\Delta\, ( -1+{m}^{2}{r_{+}}^{2} )
 ) }{{r_{+}}^{2-z} ( y_{{m}}-1 )  (  ( \Delta-2
 ) y_{{m}}-\Delta ) }}}
 .
\end{eqnarray}
By using the density (\ref{davood}), one has that
$\langle{\cal O}_+\rangle$
can be expressed as
\begin{eqnarray}
\langle{\cal O}_{+}\rangle
&=&-{\frac {{y_{{m}}}^{1-\Delta} ( {r_{+}}^{2} ( -1+y_{{m}}
 ) {m}^{2}+2\,{\it } ) {r_{+}}^{\Delta}}{
 (  ( \Delta-2 ) y_{{m}}-\Delta ) r_{+}{\it }}}\sqrt {\Gamma },\label{OO}
\end{eqnarray}
where the new function $\Gamma$ is defined as
\begin{eqnarray}
\hspace{-5mm}\Gamma= {\frac { \Big( -\sqrt {2} (  ( \frac{z}{2}+N-4 ) y
_{{m}}-\frac{z}{2}-N+5 ) \Sigma\,{r_{+}}^{\frac{z}{2}+N-3}+ ( \frac{z}{2}+N-3
 ) {y_{{m}}}^{\frac{z}{2}+N-4}\rho \Big) \sqrt {2}{\it B'(1)}}{\Sigma
\,{r_{+}}^{\frac{z}{2}+N-3} ( 1-y_{{m}} ) }}\,.
\end{eqnarray}
Now we can write $\Gamma$ in the following equivalent form
\begin{eqnarray}
\Gamma&=& A\,\frac{T_c-T_H}{\sqrt{T_H}},
\end{eqnarray}
where $T_H$ is
the Kodama temperature (\ref{H})
and $A$ is a function of $\{r_{+},\Delta,y_m\}$.
% The critical temperature $T_c$ is defined  as
%\begin{eqnarray}
%\frac{T_c}{\sqrt{\rho}}&=&\frac{( \frac{z}{2}+N-3
 %) {y_{{m}}}^{\frac{z}{2}+N-3}}{4\sqrt {2}y_{{m}}\pi r_+^{\frac{z}{2}+N-2}( -( \frac{z}{2}+N-4 ) y
%_{{m}}+\frac{z}{2}+N-5 )}\times\nonumber\\
%&&\hspace{5cm}\sqrt {\frac {{r_{+}}^{2-z} (1- y_{{m}} )  \Big(( \Delta-2
 %) y_{{m}}-\Delta \Big)}{ -{m}^{2}{r_{+}}^{2} ( \Delta-1
 %) y_{{m}}-\Delta\, ( 1-{m}^{2}{r_{+}}^{2} )
  %}}\label{Tc}\,.
%\end{eqnarray}
The expression in (\ref{OO}) is complicated and it is hard to extrapolate the critical temperature from it. Furthermore, the values of the parameters depend on the matching point $y_m$ in the bulk and this equation is not well written in terms of our physical parameters $\rho, T_H$ instead the horizon radius $r_+$. In what follows, we will reduce and examine these expression in some simple but physically important cases. The mass of the scalar field can be set to zero, without loss of the generality. In addition, the location of the radial matching point is fixed at $y_m=\frac{1}{2}$. As the first step, we will furnish the expression of $r_+$ in terms of $T_H,l_{eff}$ from (\ref{r+}). However, the derived expression still remains so complicated. For this reason, we will discuss the cases of different values of $z,N,\tilde{k}$ separately. In any case, we will write the critical temperature $T_c$ as a function of $\rho$. For the sake of simplicity, we also fix $V_0=1,l_{eff}=1$.

%%%%%%%%%%%%%%%%%%%%%%%%%%%%%%%%%%%%%%%%%%%%%%%%%%%%%%%%%%%%%
\subsection{The Lorentz invariance case in $N=4$: $z=0$}
%%%%%%%%%%%%%%%%%%%%%%%%%%%%%%%%%%%%%%%%%%%%%%%%%%%%%%%%%%%%

By setting $z=0$, $N=4$ and $k=0$ (planar case) in (\ref{OO}), we get 
\begin{eqnarray}
\langle{\cal O}_{+}\rangle
&\sim&0.0063304(12.566T_H+\sqrt {157.92T_H^2- 6}) ^
2\\&& \nonumber\times\sqrt{5189.8T_H+ 412.97(157.92T_H^2-6)^{1/2}+90
\rho}\label{O1}
\end{eqnarray}
We have that this scalar oparator vanishes at
\begin{equation}
T_c=-277.9132603\rho+0.0009828667268\sqrt{7.994700000\times10^{10}\rho^2-6.30253302\times10^8}\label{Tc1}\,,
\end{equation}
such that we can expand (\ref{O1}) near $T_c$ as 
\begin{eqnarray}
\langle{\cal O}_{+}\rangle
&\sim&(T_c-T_H)^{1/2}\,.
\end{eqnarray}
It is very interesting to observe that in such a case, in order to have a real critical temperature, $\rho\geq 0.0887$. It means there exists a lower bound on $\rho$ in which below it no condensation happens.

For $k=-1$ we obtain
\begin{eqnarray}
\langle{\cal O}_{+}\rangle
&\sim&
0.0063304(12.566T_H+\sqrt{157.92T_H^2+6}) ^
2\\&& \nonumber\times\sqrt{5189.8T_H+412.97\sqrt{157.92T_H^2+6}+90
\rho}\,,\label{O2}
\end{eqnarray}
such that
\begin{eqnarray}
T_c=-277.9132603\rho+0.0009828667268\sqrt{7.994700000\times10^{10}\rho^2+6.30253302\times10^8}\label{Tc2}\,.
\end{eqnarray}
We observe that the behavior of (\ref{O2}) near the critical temperature is the same of (\ref{O1}), but now we do not recover a minimal value for $\rho$.

Finally, for $k=1$, we get
\begin{eqnarray}
\langle{\cal O}_{+}\rangle
&\sim&
0.025321\, \left(  6.2832\,T_H+\sqrt { 39.479\,{T_H}^{2}- 3.0} \right) ^{
2}\\&& \nonumber\times\sqrt { 5189.7\,T_H+ 825.98\,\sqrt { 39.479\,{T_H}^{2}- 3.0}+ 90.0\,\rho
}\,,\\
&&\hspace{-20mm}T_c=363.8366637\rho+0.0001286830142\sqrt{7.994497500\times10^{12}\rho^2+9.628077237\times10^{10}}\,.
\end{eqnarray}
Also in this case  $\langle{\cal O}_{+}\rangle\sim(T_c-T_H)^{1/2}$ and the critical temperature behaves as $T_c\sim \sqrt{\rho}$, according with the literature about s-wave holographic superconductors and it increases monotically. This behavior will change only in the presence of the higher order corrected backgrounds like Weyl's models for s-wave. For example, for Gauss-Bonnet and Weyl corrections, it reads as  $T_c=\sqrt[3]{\rho}$ \cite{gregory}.

%%%%%%%%%%%%%%%%%%%%%%%%%%%%%%%%%%%%%%%%%%%%%%%%%%%%%%%%%%%%%
\subsection{Lorentz invariance  breaking in $N=4$: $z=1$}
%%%%%%%%%%%%%%%%%%%%%%%%%%%%%%%%%%%%%%%%%%%%%%%%%%%%%%%%%%%%5

Any Lifshitz redshift paramter $z\neq0$ breaks the Lorentz symmetry by breaking the footing of the space and time coordinates. In this Subsection, in order to clearify the effect of the $z$ on condensation, we consider the case of $z=1$ in $N=4$, in order to recover a more realistic three dimensional holographic superconductor in the absence of the Lorentz symmetry.

We start by the case $k=0$ and we get
\begin{eqnarray}
 \langle{\cal O}_{+}\rangle
&\sim&0.0076809(12.566\,T_H+\sqrt { 157.92\,{T_H}^{2}- 7}) ^
{2}\\&& \nonumber\times\sqrt {{\frac { 30743\,{T_H}^{2}+ 2446.5\,T_H\sqrt { 157.92\,{T_H}^{2}-
 7}- 681.41+ 245\,\rho}{ 12.566\,T_H+\sqrt { 157.92\,{T_H}^{2}- 7}}}
}\,,
\end{eqnarray}
and 
\begin{eqnarray}
T_c&=&860190000000+{ 3.5945\times 10^{16}}\,\rho\\&& \nonumber +23351000000\,\sqrt {
 2369800000000\,{\rho}^{2}- 951480000\,\rho+ 1480900000}\,.
\end{eqnarray}
In this case $\rho$ is unbounded, since the argument of the root always is positive.

For $k=-1$ we have
\begin{eqnarray}
 \langle{\cal O}_{+}\rangle
&\sim&0.0076809\, \left(  12.566\,T_H+\sqrt { 157.92\,{T_H}^{2}+ 7} \right) ^
{3/2}\\&& \nonumber\times\sqrt { 30743\,{T_H}^{2}+ 2446.5\,T_H\sqrt { 157.92\,{T_H}^{2}+ 7}+
 681.41+ 245\,\rho}\,,
\end{eqnarray}
and the critical temperature reads
\begin{eqnarray}
T_c&=&0.5238514818\times10^{-7}\,\sqrt {- 8.6\times10^{11}+ 35945469570000000.0\,
\rho+ 23351084080.0\eta}\,,\nonumber\\ \\
\eta&=&\,\sqrt { 2369787000000.0\,{\rho}^{2}+ 951482000.0
\,\rho+ 1480876313.0}\,.
\end{eqnarray}
Finally, for $k=1$, we get
\begin{eqnarray}
 \langle{\cal O}_{+}\rangle
&\sim& 0.0076809\, \left(  12.566\,T_H+\sqrt { 157.92\,{T_H}^{2}- 21} \right) 
^{2}\\&& \nonumber\sqrt {{\frac { 30743.0\,{T_H}^{2}+ 2446.5\,T_H\sqrt { 157.92\,{T_H}^{2}
- 21}- 2044.2+ 245\,\rho}{ 12.566\,T_H+\sqrt { 157.92\,{T_H}^{2}- 21
}}}}\,,\\
T_c&=&0.02619257409\times10^{-6}\,\sqrt { 279282401700+ 1437818783000000\,\rho+\zeta}\,,\\
\zeta&=&
 4670216817\,\sqrt { 94791480000\,{\rho}^{2}- 90963600\,\rho+
 533102913}\,.
\end{eqnarray}
We can see that, in all the topological cases, $\langle{\cal O}_{+}\rangle$ is zero at a specific value of $T_H=T_c$,
namely the critical point, and
condensation occurs at $T_H<T_c$. The important point is that the values of the critical temperature and the condensation scheme depend on the toplogical parameter $k$ and on the Lifshitz scaling $z$. Furthermore, we showed that the behavior of 
$\langle{\cal O}_{+}\rangle \propto(1-T_H/T_c)^{1/2}$
always is recovered, in agreement with the literature. The system below this critical temperature becomes superconductor, but the critical exponent of the model remains the same of the usual holographic superconductor without the higher order gravitational corrections.

\section{Discussions}
In this paper we have considered a holographic model for a non-relativistic system showing
superconductivity. We have used a black hole background which comes from  a scalar field model  minimally coupled
with the Abelian gauge field in the presence of cosmological
constant $\Lambda$ in N dimensions, and we have studied analytically holographic superconductors in this new
kind of asymptotic AdS solutions. We have considered static, (pseudo-)spherically symmetric (SSS) solutions with various topologies in two different cases, $\Lambda=0$ and $\Lambda\neq0$. We have obtained the quasi-local generalized Misner-Sharp mass as a Killing conserved charge. This quasi-local energy at the horizon $r=r_{+}$ is identified with black hole energy. Then we have derived the Wlad entropy, Killing-Hawking temperature and Kodama-Hayward temperature of black hole solutions. These temperatures are in principle different. 
We have shown that for our non-vacuum solution, the first law of black hole thermodynamics holds true by making use of Kodama temperature:
this argument substantiates our proposed temperature which is different with respect to the Hawking one generally chosen
in literature.
After that we have studied the hairy black hole solutions in which near the horizon the Abelian gauge field breaks the symmetry. In the holographic picture, this symmetry breaking is equivalent to a second order phase transition near the horizon. We also have analytically solved the system in
the probe limits, near horizon and asymptotic region. We have found that there is also a critical temperature which is a function of the Lifshitz parameter $z$ and under such temperature a condensation field
appears. 
The value of this critical temperature also depends on the topology. In some special topological cases, the critical temperature appears only for $\rho\in (\rho^*, \infty)$, where $\rho^*$ is a fixed (minimal) value of the charge density, so that $T_c(\rho^*)$ corresponds to the best configuration for the holographic supercoductor.

\section*{Acknowledgments}
   We would like to thank Dr. Yanyan Bu  and Prof. R. B. Mann. L.S. thanks also Prof. S. Zerbini for valuable suggestions. 

\section*{Appendix A. The Kodama/Hayward temperature}

In this Appendix, we would like to give a brief derivation of the Kodama/Hayward temperature.
In four dimensions, 
any spherically symmetric (dynamical) metric can be expressed in the following form,   
\begin{equation*}
\label{metric1}
ds^2 =\gamma_{ij}(x^i)dx^idx^j+ {\mathcal R}^2(x^i) d\Omega_2^2\,,\qquad i,j \in \{0,1\}\;,
\end{equation*} 
where the two-dimensional metric
\begin{equation*} 
d\gamma^2=\gamma_{ij}(x^i)dx^idx^j\,,
\end{equation*}
is referred to as the normal one with the related coordinates $\{x^i\}$, while
$ {\mathcal R}(x^i)$ is the areal radius, considered as a scalar field in the two dimensional
normal space. 
A relevant scalar quantity in the reduced normal space is given by
\begin{equation*}
\chi(x^i)=\gamma^{ij}(x^i)\partial_i  {\mathcal R}(x^i)\partial_j  {\mathcal R}(x^i)\,, \label{sh} 
\end{equation*} 
since the dynamical trapping horizon, if exists, is located in
correspondence of 
\begin{equation*} 
\chi(x^i)\Big\vert_H = 0\,,\quad \partial_i\chi(x^i)\vert_H \gneq 0\,.\label{ho} 
\end{equation*}
In the static case, if we refer to the metric (\ref{metric}), we have $\chi(r)=B(r)$ and in General Relativity,
by means of the time-like Killing vector field $K^\mu=\left(1,0,0,0\right)$, 
we have that the Misner Sharp mass corresponds to the charge of the conserved current $J_{\mu}=G_{\mu\nu}K^\nu$, where $G_{\mu\nu}$ is the Einstein tensor.
In the dynamical case, where we do not have the time-like Killing vector field, in order to define a conserved current, we need the Kodama vector field~\cite{Kodama}
\begin{equation*} 
\mathcal K^i(x^i):=\frac{1}{ \sqrt{-\gamma}}\,\varepsilon^{ij}\partial_j{\mathcal R}(x^i)\,,\,\,i=0,1\,;
\qquad \mathcal K^i:=0\,,\,\,i\neq 0,1\,,
\end{equation*} 
where $\varepsilon^{ij}$ is the completely antisymmetric Levi-Civita tensor on the normal 
space and $\gamma$ the determinant of $\gamma_{ij}$ metric tensor. Thus, 
the Kodama/Hayward surface gravity associated with dynamical
horizon is given by the normal-space scalar 
\begin{equation*}
\kappa_H:=\frac{1}{2}\Box_{\gamma} {\mathcal R(x^i)}\Big\vert_H\,, \label{H} 
\end{equation*} 
where $\Box_\gamma$ is the Laplacian corresponding to the $\gamma_{ij}$ metric.
In Ref. \cite{H}, Hayward showed that on the dynamical, trapping horizon the following identity holds true
\begin{equation*}
\Delta E=\frac{\kappa_{H}}{2 \pi} \Delta S
-p d\mathcal V_H\,,
\end{equation*}
where $E$ is the Misner Shiarp mass of the black hole, $S$ the entropy and $p$ the matter pressure. Thus, in order to find the Gibbs relation of thermodynamic, we define the Hayward/Kodama temperature as
\begin{equation*} 
T_H=\frac{\kappa_H}{2\pi}\,.
\end{equation*}
In the static case, it simply results to be $T_H=B'(r_H)/4\pi$.

\end{document}